# A tool for the morphological classification of galaxies: the concentration index

Xin-Fa Deng

School of Science, Nanchang University, Jiangxi, China, 330031

**Abstract** Using the galaxy data of the Sloan Digital Sky Survey Data Release 8 (SDSS DR8), I explore whether the concentration index is a good morphological classification tool and find that a reasonably pure late-type galaxy sample can be constructed with the choice of r-band concentration index ci=2.85. The opposite is not true, however, due to the fairly high contamination of an early-type sample by late-type galaxies. In such an analysis, the influence of selection effects is less important. To disentangle correlations of the morphology and concentration index with stellar mass, star formation rate (SFR), specific star formation rate (SSFR) and active galactic nucleus (AGN) activity, I investigate correlations of the concentration index with these properties at a fixed morphology and correlations of the morphology with these properties at a fixed concentration index. It is found that at a fixed morphology, high-concentration galaxies are preferentially more massive and have a lower SFR and SSFR than low-concentration galaxies, while at a fixed concentration index, elliptical galaxies are preferentially more massive and have a lower SFR and SSFR than spiral galaxies. This result shows that the stellar mass, SFR and SSFR of a galaxy are correlated with its concentration index as well as its morphology. In addition, I note that AGNs are preferentially found in more concentrated galaxies only in the spiral galaxy sample.

## 1. Introduction

The study of galaxy morphologies has long been an important issue. Numerous authors have explored the environmental dependence of galaxy morphologies (e.g., Postman & Geller 1984; Dressler et al. 1997; Hashimoto & Oemler 1999; Fasano et al. 2000; Tran et al. 2001; Goto et al. 2003; Helsdon & Ponman 2003; Treu et al. 2003; Deng et al. 2007a-c, 2008a-b, 2009a), and it is widely believed that early-type galaxies tend to reside in the densest regions of the universe, while late-type galaxies tend to reside in low density regions. There have also been numerous works that focus on correlations between galaxy morphologies and other parameters. For example, some studies have shown that high-luminosity galaxies are preferentially ''early type'' (e.g., Blanton et al. 2003; Baldry et al. 2004; Balogh et al. 2004; Kelm et al. 2005). Another typical correlation between galaxy morphologies and other parameters is the correlation between galaxy morphologies and colors (e.g., Holmberg 1958; Roberts & Haynes 1994; Strateva et al. 2001). Strateva et al. (2001) indicated that blue galaxies are dominated by late types, while red galaxies are dominated by early types. Many works also shed light on galaxies with a given morphological type, especially early-type galaxies. Thomas et al. (2002) found that the $\alpha$/Fe-$\sigma$ relation (and its scatter) of early-type galaxies is independent of the environmental density. Moss & Whittle (2005) concluded that the frequency of emission-line galaxies is similar for field and cluster early-type galaxies. Deng et al. (2009b) and Deng (2010) studied the environmental dependence of luminosity, g-r color, star formation rate (SFR) and specific star formation rate (SSFR) at a fixed morphology. Tempel et al. (2011) investigated the galaxy luminosity function of different morphological types at different environmental density levels.

Undoubtedly, the key question of the above-mentioned works is how to morphologically classify galaxies. The traditional method of morphological classification is to visually inspect galaxy images according to Hubble's classification scheme (Sandage 1961). Because it is highly labor intensive, however, the visual inspection procedure limits the size of galaxy samples. Thus, it is desirable to find automated morphological classification schemes to classify large numbers of galaxies into early and late types. There are a number of parameters, such as concentration index, color, spectral features, surface brightness profile, structural parameters or some combination of these, that exhibit a strong correlation with morphological type and can be used to classify galaxies (e.g., Shimasaku et al. 2001; Strateva et al. 2001; Abraham, van den Bergh & Nair 2003; Nakamura et al. 2003; Kauffmann et al. 2004; Park & Choi 2005; Yamauchi et al. 2005; Conselice 2006; Sorrentino et al. 2006; Scarlata et al. 2007).

The concentration index is known to correlate with the morphological type (Morgan 1958; Doi et al. 1993; Abraham et al. 1994; Shimasaku et al. 2001; Nakamura et al. 2003; Park & Choi 2005). Shimasaku et al. (2001) showed that the (inverse) concentration index $C = r_{50} / r_{90}$, defined as the ratio of the half-light Petrosian radius to the 90% light Petrosian radius, is closely correlated with the morphological type. This index is useful for the automated classification of early- and late-type galaxies if one is satisfied with a completeness of $\approx$ 70%-90%, allowing for a contamination of $\approx$ 15%-20%. Shimasaku et al. (2001) also examined the correlations of visual morphology with a number of parameters measured by PHOTO. They found that the concentration index shows the strongest correlation with visual morphology and that a combination of the concentration index with other parameters, such as surface brightness, color and asymmetry, does not appreciably enhance the correlation. They therefore concluded that the concentration index is perhaps the best parameter for classifying galaxy morphology, which is consistent with the conclusion of Doi et al. (1993) and Abraham et al. (1994). Nakamura et al. (2003) separated galaxies into early and late types according to C < 0.35 and C > 0.35, which corresponds to a division at S0/a. When the visually classified sample is taken as the reference, the early-type and late-type galaxy samples classified by Nakamura et al. (2003) show an 82% completeness and an 18% contamination from the opposite sample. Park & Choi (2005) used the color versus color gradient space as the major morphological classification tool and the concentration index as an auxiliary parameter. They found early-type galaxies to be strongly concentrated within a spot centered at (2.82,−0.04) in the u−r and $\Delta$(g −i) planes, with the center at (2.82, 0.3) in the u-r color-concentration index space. The upper panel of Fig.1 of Park & Choi (2005) shows that the concentration index is still a relatively good and simple parameter for classifying the morphology of galaxies, and most early-type galaxies have a concentration index $c = R_{50} / R_{90} < 0.35$.

Galaxy Zoo is a web-based project (http://www.galaxyzoo.org) that uses the collective efforts of hundreds of thousands of volunteers to produce morphological classifications of galaxies (Lintott et al. 2008, 2011). Lintott et al. (2008) found that there is a remarkable degree of agreement (better than 90% in most cases) between this data set and those compiled by professional astronomers, thus demonstrating that the data from volunteers provide a robust morphological catalog. Each object in this project is classified as belonging to one of six categories: Spiral (clockwise rotation), Spiral (anticlockwise rotation), Spiral (edge-on/rotation unclear), Elliptical, Merger, or Star/Don't Know. All three possible spiral classifications are often combined into a single classification, which is useful for studies that require only a simple split into elliptical and spiral samples. Full details on the classification process, including the operation of the site, are given in Lintott et al. (2008). Table 2, compiled by Lintott et al. (2011), has recently been made public. The table contains the data for all Main galaxies (MGS, Strauss et al. 2002) in the Sloan Digital Sky Survey Data Release 7 (SDSS DR7) (Abazajian et al. 2009). This table includes the raw votes, the weighted votes in elliptical (E) and combined spiral (CS=clockwise +anticlockwise+edge-on spiral) categories and flags indicating the inclusion of the galaxy in a clean, debiased catalog.

When exploring whether the concentration index is a good morphological classification tool, previous authors only used the visual inspection samples compiled by professional astronomers. Because the visual inspection procedure is very labor intensive, such samples are fairly small, making it difficult to make statistically sound comparisons. Table 2 of Lintott et al. (2011) contains visual morphological classifications of 667,945 MGS, which is much larger than the samples used in previous studies. In this study, I attempt to use the data set from volunteers and to further investigate whether the concentration index is a good morphological classification tool.

This paper is organized as follows. In section 2, I describe the data used. In sections 3 and 4, I examine whether the concentration index is a good morphological classification tool, and I discuss the statistical results. My main results and conclusions are summarized in section 5.

## 2. Data

A new program called SDSS-III began operation in August 2008 and will continue through July 2014. Using the SDSS facilities at the Apache Point Observatory (APO), SDSS-III will carry out four surveys: SEGUE-2, the Baryon Oscillation Spectroscopic Survey (BOSS), the Multi-object APO Radial Velocity Exoplanet Large-area Survey (MARVELS) and the Apache Point Observatory Galactic Evolution Experiment (APOGEE). Eisenstein et al. (2011) has described this program in detail.

As with SDSS-I/II, SDSS-III data will periodically be released publicly. The first of these releases, referred to as the eighth data release(DR8) (Aihara et al. 2011), is already available. As with previous data releases, DR8 is cumulative and includes essentially all data from the previous releases. Aihara et al. (2011) claimed that DR8 is not just a repeat of previous data releases but is also an enhancement. SDSS-III has reprocessed all SDSS-I/II imaging data and all stellar spectra.

Fortunately, the data of the initial Galaxy Zoo classifications are included in DR8. DR8 also contains a number of physical galaxy parameters derived by the MPA-JHU group, such as BPT classification, stellar mass, nebular oxygen abundance, star formation rates (SFRs) and the specific SFR (SSFR). In future works, I will download these important parameters from the Catalog Archive Server of SDSS Data Release 8 (Aihara et al. 2011).

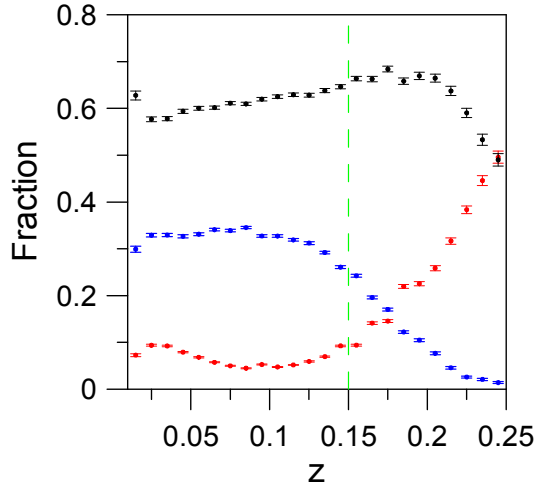

Fig.1 Galaxy Zoo type fractions as a function of redshift: red, blue and black dots represent elliptical, spiral and uncertain galaxies, respectively. The error bars of the blue lines are 1 $\sigma$ Poissonian errors. Selection effects can be removed below z=0.15 (green vertical dashed line).

In this work, I downloaded the initial Galaxy Zoo classifications and other parameters of the Main galaxy sample (Strauss et al. 2002) using the SDSS SQL Search (with SDSS flag: bestPrimtarget&64>0) with a redshift range of 0.01<z<0.25. In this catalog, those galaxies whose debiased votes give an unambiguous answer (> 80%) of their morphology are explicitly labeled as elliptical or spiral, and all other galaxies are flagged as uncertain. My sample contains 617672 Main galaxies: 55112 elliptical, 178557 spiral and 384003 uncertain. Fig. 1 shows the Galaxy Zoo type fractions as a function of redshift. Bamford et al. (2009) indicated that assuming that there is negligible evolution in the galaxy population over the redshift interval considered and that the survey contains a similar distribution of environments at each redshift, then the true type fractions should be constant with the redshift. Any trends in the observed GZ type fractions with redshift may be attributed to a combination of two biases: selection and classification. The selection bias is due to the variation in the size and luminosity distribution of galaxies in the magnitude-limited sample. In the classification bias, for otherwise morphologically identical galaxies, objects that are apparently fainter and smaller are more likely to be classified as early-type due to the diminished spatial resolution and the signal-to-noise ratio. Bamford et al. (2009) statistically corrected for this classification bias. In this work, I use flags indicating the inclusion of a galaxy in a clean, debiased catalog. As seen in Fig.1, in the redshift range z<0.15, the de-biased type fractions are approximately flat. I believe that this redshift range, in which there are 31202 elliptical galaxies and 160577 spiral galaxies, is free from selection effects.

## 3. Correlation of the concentration index with the Galaxy Zoo types

$R_{50}$ and $R_{90}$ are the radii enclosing 50% and 90% of the Petrosian flux, respectively. In this study, I calculate the concentration index of r-band ci = $R_{90}/R_{50}$. Fig.2 shows the ci distributions of elliptical and spiral galaxies for a full sample and a sample with a redshift range of z<0.15 in which selection effects are not present. Indeed, the bimodality of the ci distribution can be observed: the majority of spiral galaxies correspond to low-concentration galaxies, while the majority of elliptical galaxies correspond to high-concentration galaxies.

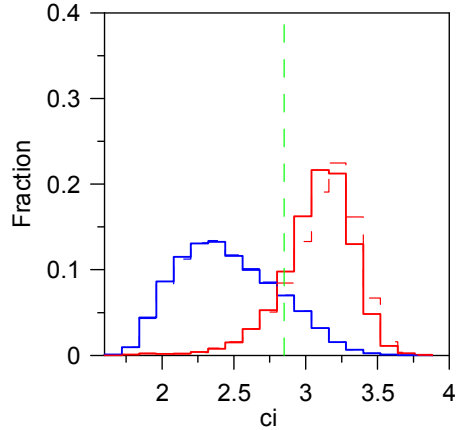

Fig.2 Ci distributions of elliptical and spiral galaxies for a full sample (solid line) and a sample with a redshift range of z<0.15 (dashed line): the red line represents elliptical galaxies and the blue line represents spiral galaxies. The green vertical dashed line indicates ci=2.85.

Like Shimasaku et al. (2001), I study the completeness and contamination of the morphologically classified sample with the use of ci. The left panel of Fig.3 shows the completeness as a function of ci: the red curve represents the completeness of the early-type galaxy sample with a concentration index larger than a given ci; the blue curve represents the completeness of the late-type galaxy sample with a concentration index smaller than a given ci. The completeness of the two samples balances with a ci=2.85 at $\approx 85\%$. In this work, I select ci=2.85 as the separator point between early-types and late-types, which is the same as that obtained by Nakamura et al. (2003). However, the right panel of Fig.3 indicates contamination from the opposite type: the red curve is the contamination from late-type galaxies to the early-type galaxy sample; the blue curve is the contamination by early-type galaxies to the late-type sample. Table 1 lists the completeness and contamination of early- and late-type galaxies at ci=2.85. As seen from Table 1 and the right panel of Fig.3, the contamination of an early-type sample by late-type galaxies is much larger than that obtained by Shimasaku et al. (2001), while the contamination of the late-type sample by early-type galaxies is much smaller. One possible explanation for these differences is that the spiral : elliptical ratio (178557/55112=3.2 in the full sample) in this work is much larger than the $T \geq 1.5$ galaxies : T<1.5 galaxies ratio (290/136=2.1) in the study of Shimasaku et al. (2001). The above statistical results show that a reasonably pure late type galaxy sample can be constructed with the choice of ci=2.85; however the opposite is not true due to the fairly high contamination of an early-type sample by late-type galaxies, which is consistent with the conclusion by Shimasaku et al. (2001). Thus, when classifying galaxies into morphological classes using the concentration index, one must treat the statistical results of the early-type sample with caution.

I also show results of the sample with a redshift range of z<0.15, which is free from selection effects. As seen from Figs.2-3 and Table 1, in such an analysis the influence of selection effects is less important.

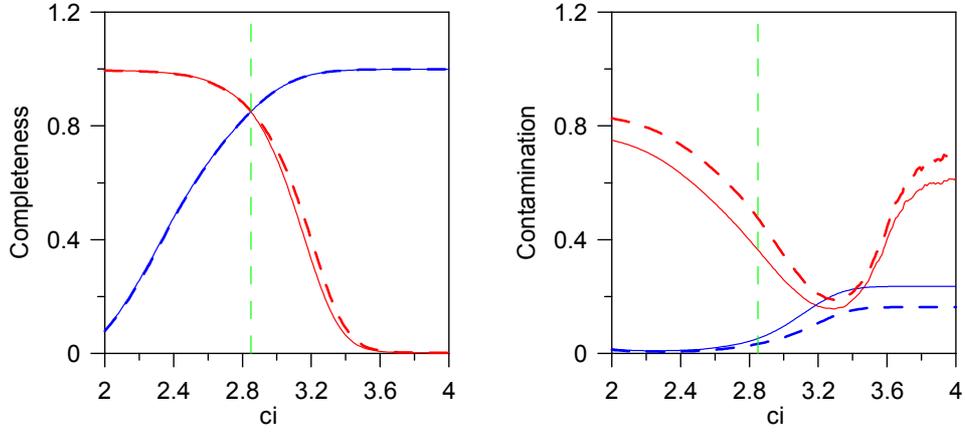

Fig.3 Completeness (left panel)and contamination (right panel) of early- and late-type galaxies as a function of ci for a full sample (solid line) and a sample with a redshift range of z<0.15 (dashed line): the red line represents early-type galaxies, and the blue line represents late-type galaxies. The green vertical dashed line indicates ci=2.85.

Table1: Correlation between visual morphology and that classified with the concentration index of r-band ci =2.85

| Sample | Parameter | Elliptical | Spiral | Sum | Contamination(%) |
|---|---|---|---|---|---|
| Full sample | "Early-type" galaxies | 46768 | 26674 | 73442 | 36.32 |
|  | "Late-type" galaxies | 8344 | 151883 | 160227 | 5.21 |
|  | Sum | 55112 | 178557 |  |  |
|  | Completeness(%) | 84.86 | 85.06 |  |  |
| Sample with a redshift range of z<0.15 | "Early-type" galaxies | 26582 | 23980 | 50562 | 47.43 |
|  | "Late-type" galaxies | 4620 | 136597 | 141217 | 3.27 |
|  | Sum | 31202 | 160577 |  |  |
|  | Completeness(%) | 85.19 | 85.07 |  |  |

Galaxy morphology and the concentration index are strongly correlated with many other properties (e.g., Holmberg 1958; Kennicutt 1992; Roberts & Haynes 1994; Strateva et al. 2001; Blanton et al. 2003; Kauffmann et al. 2003a, 2003b; Baldry et al. 2004; Balogh et al. 2004; Kelm et al. 2005; Bamford et al. 2009; Deng 2010; Deng et al. 2010). For example, it is widely accepted that early-type galaxies are redder and more luminous. Kennicutt (1992) found that morphological type is strongly correlated with the star formation rate (SFR). Bamford et al. (2009) also showed that galaxy morphology is sensitive to stellar mass. Kauffmann et al. (2003a) demonstrated a sharp transition in the physical properties of galaxies at a stellar mass of $\approx 3 \times 10^{10} M_\Theta$ and that low-mass galaxies have low concentration indices typical of disks, while high-mass galaxies have high concentration indices typical of bulges. Deng et al. (2010) noted the correlation between star formation activities and the concentration index: passive galaxies are more luminous, redder, highly concentrated and preferentially "early-type". As seen from Figs.3-6 of Deng (2010), highly concentrated or "early-type" galaxies preferentially have a lower star formation rate (SFR) and specific star formation rate (SSFR).

Due to the correlation between morphology and the concentration index, strong correlations between a parameter and other galaxy properties will result in correlations between the another parameter and these galaxy properties. To disentangle correlations of morphology and the concentration index with stellar mass, SFR and SSFR, I attempt to explore correlations of the concentration index with these properties at a fixed morphology and correlations of morphology with these properties at a fixed concentration index. Because the influence of selection effects is less important in this work, I only analyze the full sample. The full elliptical and spiral galaxy samples are divided into two subsamples each at ci=2.85. The low-concentration elliptical subsample (ci<2.85) contains 8344 galaxies, while the high-concentration elliptical subsample (ci$\geq$2.85) contains 46768 galaxies. The low-concentration spiral subsample (ci<2.85) contains 151883 galaxies, while the high-concentration spiral subsample (ci $\geq$ 2.85) contains 26674 galaxies.

The specific star formation rate (SSFR) is defined as the star formation rate (SFR) per unit stellar mass. I downloaded the total masses, total SFR and total specific SFR (SSFR) from the Catalog Archive Server of the SDSS DR8. These parameters are derived by the technique discussed in Brinchmann et al. (2004). In this study, the MEDIAN estimate is used. Figs.4, 6 and 8 show stellar mass, SFR and SSFR distributions for the low- and high-concentration elliptical galaxies(left panel) and the low- and high-concentration spiral galaxies(right panel). As seen from these figures, at a fixed morphology, the high-concentration galaxies are preferentially more massive and have a lower SFR and SSFR than the low-concentration galaxies. Figs.5, 7 and 9 also illustrate stellar mass, SFR and SSFR distributions for low-concentration elliptical and spiral galaxies (left panel) and high-concentration elliptical and spiral galaxies (right panel). In these figures, I note that at a fixed concentration index, elliptical galaxies are preferentially more massive and have a lower SFR and SSFR than spiral galaxies. All these results show that the stellar mass, SFR and SSFR of a galaxy are correlated with its concentration index as well as its morphology.

To reach a statistically sound conclusion, I also performed the Kolmogorov-Smirnov (KS) test, which checks whether two independent distributions are similar or different by calculating a probability value. The lower the probability value is, the less likely the two distributions are similar. Conversely, the higher or closer to 1 the value is, the more similar the two distributions are. The probability of the two distributions in Figs.4-9 coming from the same parent distribution is nearly 0, which shows that two independent distributions completely differ in these figures. Thus, the above-mentioned statistical conclusion is robust.

I also downloaded BPT classification from the Catalog Archive Server of the SDSS DR8, then calculated the fraction of active galactic nuclei (AGNs) in each subsample: 3.99±0.22% for low-concentration elliptical galaxies, 3.56±0.09% for high-concentration elliptical galaxies, 3.43±0.05% for low-concentration spiral galaxies, and 6.19±0.15% for high-concentration spiral galaxies. Kauffmann et al. (2003b) indicated that AGNs are preferentially found in more concentrated galaxies. In this work, however, this trend is only observed in spiral galaxies.

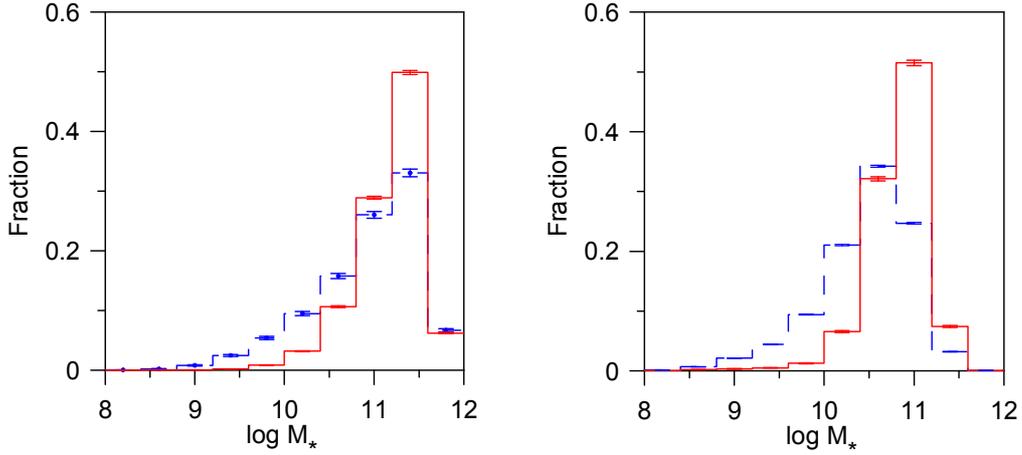

Fig.4 Stellar mass distributions for low- and high-concentration elliptical galaxies(left panel) and low- and high-concentration spiral galaxies(right panel): red solid line represents high-concentration galaxies, blue dashed line represents low-concentration galaxies. The error bars of the blue lines are 1 $\sigma$ Poissonian errors.

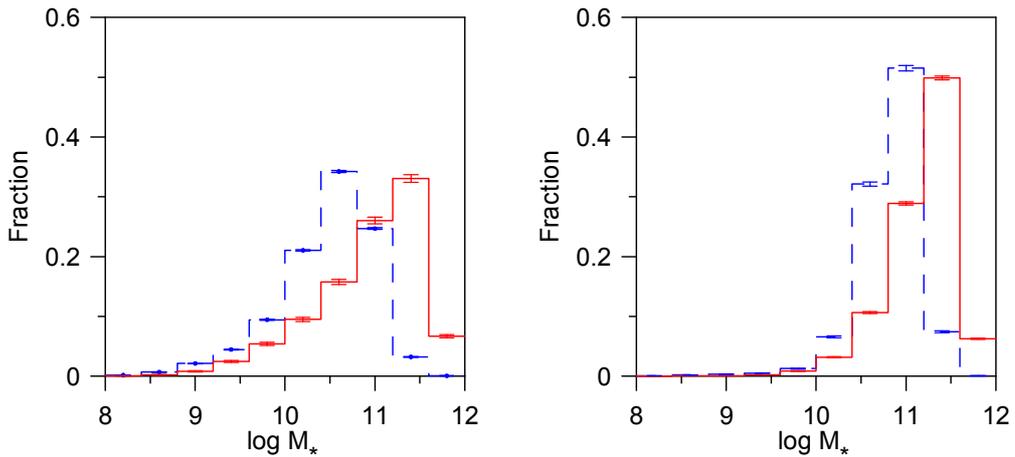

Fig.5 Stellar mass distributions for low-concentration elliptical and spiral galaxies(left panel) and high-concentration elliptical and spiral galaxies(right panel): red solid line represents elliptical galaxies, blue dashed line represents spiral galaxies. The error bars of the blue lines are 1 $\sigma$ Poissonian errors.

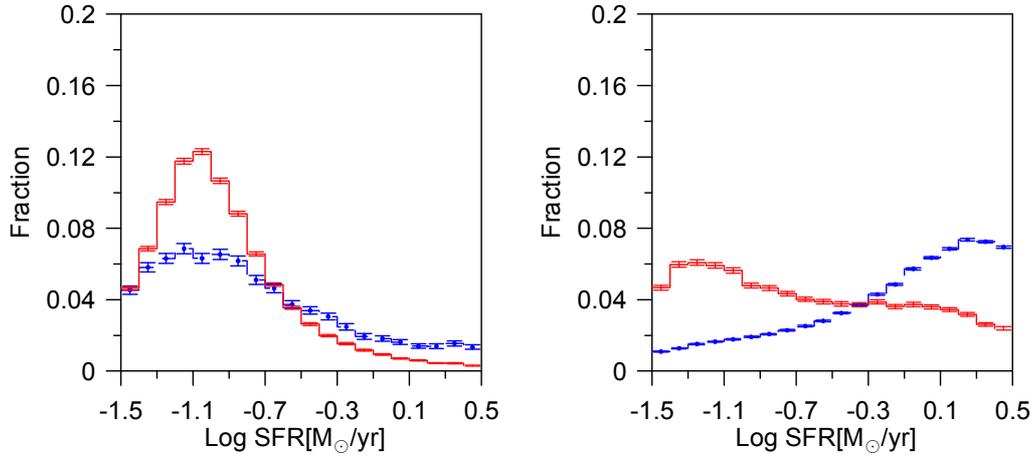

Fig.6 SFR distributions for low- and high-concentration elliptical galaxies(left panel) and low- and high-concentration spiral galaxies(right panel): red solid line represents high-concentration galaxies, blue dashed line represents low-concentration galaxies. The error bars of the blue lines are 1 $\sigma$ Poissonian errors.

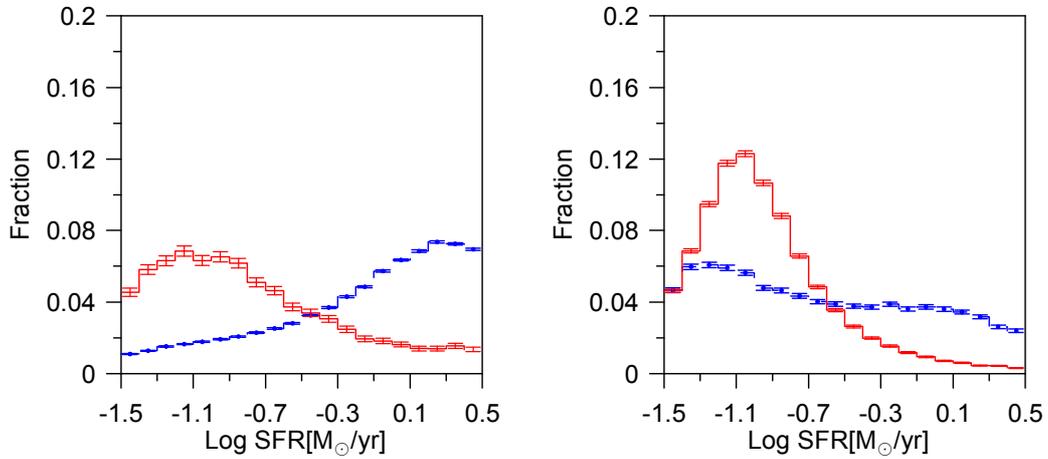

Fig.7 SFR distributions for low-concentration elliptical and spiral galaxies(left panel) and high-concentration elliptical and spiral galaxies(right panel): red solid line represents elliptical galaxies, blue dashed line represents spiral galaxies. The error bars of the blue lines are 1 $\sigma$ Poissonian errors.

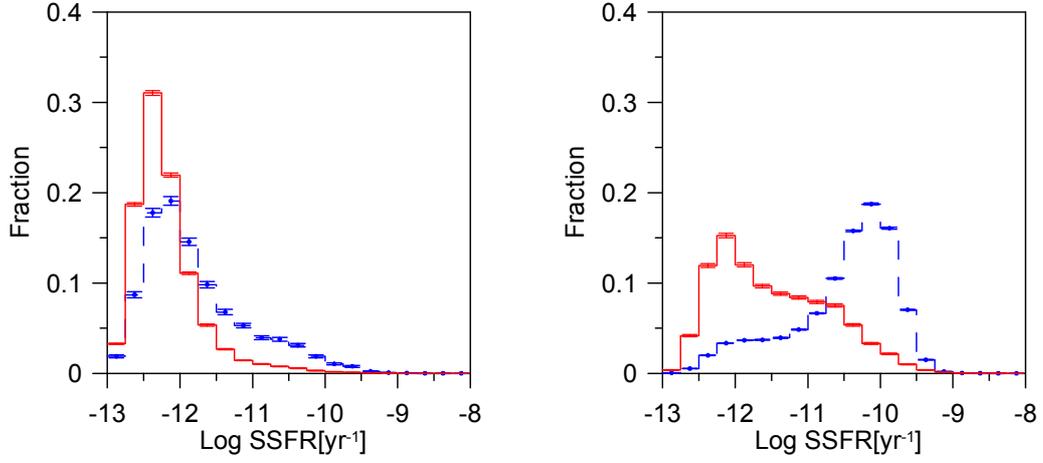

Fig.8 SSFR distributions for low- and high-concentration elliptical galaxies(left panel) and low- and high-concentration spiral galaxies(right panel): red solid line represents high-concentration galaxies, blue dashed line represents low-concentration galaxies. The error bars of the blue lines are 1 $\sigma$ Poissonian errors.

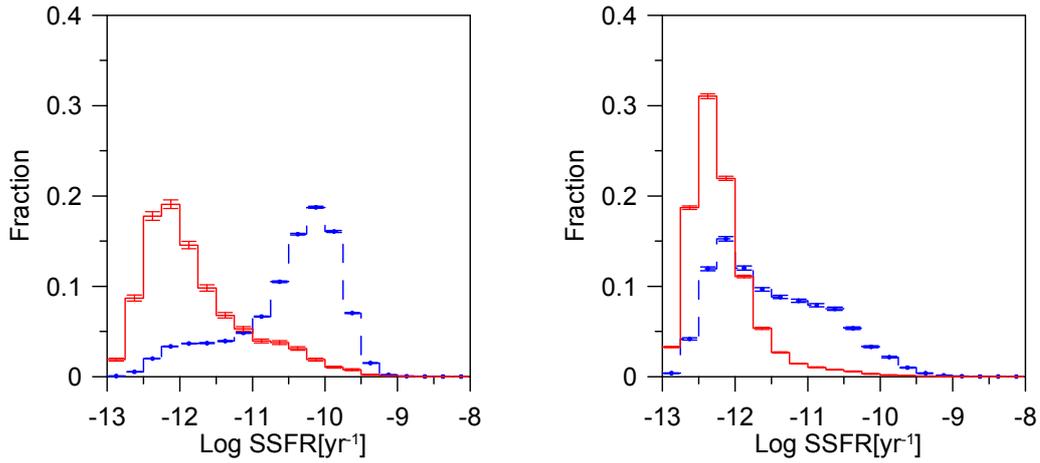

Fig.9 SSFR distributions for low-concentration elliptical and spiral galaxies(left panel) and high-concentration elliptical and spiral galaxies(right panel): red solid line represents elliptical galaxies, blue dashed line represents spiral galaxies. The error bars of the blue lines are 1 $\sigma$ Poissonian errors.

## 4. Discussion

Galaxy Zoo is a morphological classification project that utilizes visual classifications made by volunteers. Here, I note that when exploring whether the concentration index is a good morphological classification tool, there indeed is a remarkable degree of agreement between the statistical results of this data set and those compiled by professional astronomers (e.g., Shimasaku et al. 2001; Nakamura et al. 2003). This result shows that such a visually classified sample can be taken as a good reference in some works. The most important merit of such a sample is that it is much larger than the visual inspection samples compiled by professional astronomers, and thus, it can foster statistically sound conclusions.

The concentration index is a widely used parameter in automated morphological classification schemes. The statistical results of this work further show that a reasonably pure late-type galaxy sample can be constructed with the choice of ci=2.85; however the opposite is not true due to the fairly high contamination of an early-type sample by late-type galaxies. Thus, when classifying galaxies using the concentration index, a late type galaxy sample is an ideal; with an early-type sample, one must remove late-type galaxies using other auxiliary morphological classification tools.

The stellar mass, SFR and SSFR of a galaxy are correlated with its concentration index as well as its morphology, which shows that the correlation between morphology and other galaxy properties and that between the concentration index and other galaxy properties are independent of each other.

Kauffmann et al. (2003b) showed that the fraction of all galaxies classified as AGN is only a relatively weak function of the concentration index ci, while the fraction of emission-line galaxies classified as AGN rises considerably with increasing ci (see Fig.5 of Kauffmann et al. 2003b). In the work of Deng et al. (2012), all galaxies with an AGN are emission-line galaxies with 4 lines S/N>3. Deng et al. (2012) argued that AGN host galaxies have preferentially higher concentration indices than the whole galaxy sample, especially in a faint volume-limited sample. In this work, the correlation between AGN activity and concentration index ci only is observed in spiral galaxies.

## 5. Summary

Using the first data releases of SDSS-III, referred to as the eighth data release(DR8), I explore whether the concentration index is a good morphological classification tool. The statistical results show that a reasonably pure late-type galaxy sample can be constructed with the choice of ci=2.85; the opposite is not true, however, due to the fairly high contamination of an early-type sample by late-type galaxies. Such a conclusion suggests that when classifying galaxies into morphological classes using the concentration index, one must treat statistical results of the early-type sample with caution. I also analyze a sample with a redshift range of z<0.15, which is free from selection effects, and find that in such an analysis, the influence of selection effects is less important.

To disentangle correlations of morphology and the concentration index with stellar mass, SFR, SSFR and AGN activity, I investigate correlations of concentration index with these properties at a fixed morphology and correlations of morphology with these properties at a fixed concentration index. As seen from Figs.4-9, at a fixed morphology, high-concentration galaxies are preferentially more massive and have a lower SFR and SSFR than low-concentration galaxies, while at a fixed concentration index, elliptical galaxies are preferentially more massive and have a lower SFR and SSFR than spiral galaxies. All these results show that the stellar mass, SFR and SSFR of a galaxy are correlated with its concentration index as well as its morphology. Kauffmann et al. (2003b) indicated that AGNs are preferentially found in more concentrated galaxies. In this work, such a trend is only observed in spiral galaxies.

**Acknowledgements**

This study was supported by the National Natural Science Foundation of China (NSFC, Grant 11263005).

Funding for SDSS-III has been provided by the Alfred P. Sloan Foundation, the Participating Institutions, the National Science Foundation, and the U.S. Department of Energy. The SDSS-III web site is http://www.sdss3.org/.

SDSS-III is managed by the Astrophysical Research Consortium for the Participating Institutions of the SDSS-III Collaboration including the University of Arizona, the Brazilian Participation Group, Brookhaven National Laboratory, University of Cambridge, University of Florida, the French Participation Group, the German Participation Group, the Instituto de Astrofisica de Canarias, the Michigan State/Notre Dame/JINA Participation Group, Johns Hopkins University, Lawrence Berkeley National Laboratory, Max Planck Institute for Astrophysics, New Mexico State University, New York University, Ohio State University, Pennsylvania State University, University of Portsmouth, Princeton University, the Spanish Participation Group, University of Tokyo, University of Utah, Vanderbilt University, University of Virginia, University of Washington, and Yale University.